\documentclass[aps,showpacs,amsmath,amssymb,superscriptaddress,unsortedaddress,
preprintnumbers,nofootinbib,10pt]{revtex4}
\usepackage[dvipdfmx]{graphicx}
\usepackage{subfigure}
\usepackage{amsmath,amssymb,ascmac,color}
\usepackage{type1cm}
\usepackage{bm}

\newcommand\beq{\begin{equation}}
\newcommand\eeq{\end{equation}}
\newcommand\beqa{\begin{eqnarray}}
\newcommand\eeqa{\end{eqnarray}}

\newcommand{\ds}[1]{#1 \hspace{-0.5em}/}  

\newcommand\btau{\mbox{\boldmath$\tau$}}
\newcommand\bgamma{\mbox{\boldmath$\gamma$}}

\newcommand\bphi{\mbox{\boldmath$\phi$}}

\date{\today}

\begin{document}
\preprint{KUNS-2661}

\title{Chiral pair fluctuations for the inhomogeneous chiral transition}

\author{Ryo Yoshiike}
\email{yoshiike@ruby.scphys.kyoto-u.ac.jp}
\affiliation{Department of Physics, Kyoto University, Kyoto 606-8502, Japan}

\author{Tong-Gyu Lee}
\email{tonggyu.lee@yukawa.kyoto-u.ac.jp}
\affiliation{Department of Physics, Kyoto University, Kyoto 606-8502, Japan}
\affiliation{Department of Natural Science, Kochi University, Kochi 780-8520, Japan}

\author{Toshitaka Tatsumi}
\email{tatsumi@ruby.scphys.kyoto-u.ac.jp}
\affiliation{Department of Physics, Kyoto University, Kyoto 606-8502, Japan}

\begin{abstract}
The effects of fluctuations are discussed around the phase boundary of the inhomogeneous chiral transition between the inhomogeneous chiral phase and the chiral-restored phase.
The particular roles of thermal and quantum fluctuations are elucidated and a continuity of their effects across the phase boundary is suggested.
In addition, it is argued that anomalies in the thermodynamic quantities should have phenomenological implications for the inhomogeneous chiral transition.
Some common features for other phase transitions, such as those from the normal to the inhomogeneous Fulde-Ferrell-Larkin-Ovchinnikov state in superconductivity, are also emphasized.
\end{abstract}



\maketitle

\section{Introduction}
\label{sec:intro}
The inhomogeneous chiral transition is one of the fascinating topics in the study of the QCD phase diagram~\cite{fuk}.
Many people have believed that there may be the chiral transition in the chemical potential ($\mu$) and temperature ($T$) plane,
while there has not yet been any direct evidence
in addition to the fact that the lattice QCD simulations do not work at finite $\mu$ due to the so-called sign problem.
The chiral transition essentially separates two phases:
one is the spontaneous symmetry breaking (SSB) phase
and the other is the symmetric (chiral-restored) phase.
Besides these two phases, recent studies have suggested the appearance of the inhomogeneous chiral phase (iCP) as another possibility  for the realization of chiral symmetry~\cite{chi,dcdw,nic}. 
iCP is then characterized by spatially modulated chiral condensates. 
The generalized order parameter consists of the scalar or pseudoscalar condensate.
There have been proposed various kinds of structures of the condensates,
and then the characteristic features of iCP, as well as the properties of the associated phase transition, have been studied within the mean-field approximation (MFA)~\cite{bub}.
The effects of the magnetic field and the topological features have also been discussed~\cite{fro,tatb,yos1,yos2}.

While so far most discussions have been restricted to the mean-field level, 
recent studies focus on not only the properties of the Nambu-Goldstone excitations in iCP, but also the stability of iCP against such fluctuations~\cite{lee,hid}.
As an important consequence, it has been shown that for one-dimensional modulations of the condensate the correlation functions of the quark-antiquark bilinear fields exhibit quasi-long-range order (QLRO) with algebraic decay at large distances at finite temperature in accord with the Landau-Peierls theorem \cite{lan,pei},
while true long-range order is realized in the usual SSB phase.
In addition, the thermal average of the quark condensate becomes zero for $T\neq 0$ due to thermal fluctuations.
These results come from the spatially anisotropic dispersion relation of the Nambu-Goldstone modes.
Note that iCP has long-range order at $T=0$, which implies that quantum fluctuations are irrelevant for the case of the $T=0$ limit.
Similar features are found in a variety of systems, such as the smectic-A phase of liquid crystals~\cite{gen}, the Fulde-Ferrel-Larkin-Ovchinikov (FFLO) state of superconductors~\cite{fflo} or superfluids~\cite{fflo2}, the Bragg-glass phase of impure superconductors~\cite{gia}, the pion-condensed phase of nuclear matter~\cite{bfg}, and so forth.
Also, the phase with QLRO is analogous to the Berezinskii-Kosterlitz-Thouless phase~\cite{bkt} in two-dimensional systems
and its experimental verification has been done in ultracold Bose~\cite{had} and Fermi gases~\cite{mur}, as well as in exciton-polariton gases~\cite{nit}.

\begin{figure*}
\centering
\includegraphics[width=2.5in]{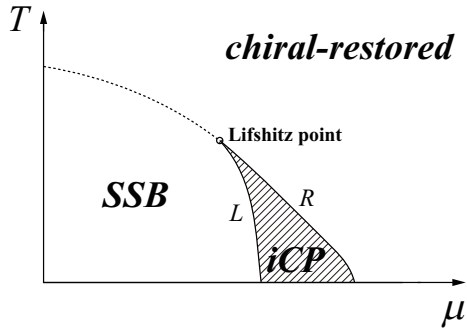}
 \caption{Schematic phase diagram for chiral symmetry breaking in the $(\mu,T)$ plane.
The shaded domain enclosed by the left- ($L$-) and right-hand-side ($R$-) boundaries represents the inhomogeneous chiral phase (iCP).
The $L$-boundary is of first or second order, depending on the type of iCP~\cite{dcdw,nic,2d}, whereas the $R$-boundary is always of second order, universally.
The boundary between the SSB and chiral-restored phases is of second order in the chiral limit (dotted line).}
 \label{fig1}
~\\
\end{figure*}

In this paper we elucidate another interesting aspect of the fluctuations near the phase boundary.
Starting from the Lifshitz point, iCP is enclosed by the two phase boundaries on the $\mu$-$T$ plane (see Fig.~\ref{fig1}): 
one is the $L$-boundary separating the usual SSB phase and iCP at lower $\mu$, 
and the other is the $R$-boundary in contact with the chiral-restored phase at larger $\mu$.
It has then been shown that the $L$-boundary has different orders and properties of the phase transition, depending on the type of the condensates~\cite{dcdw,nic,2d},
while the $R$-boundary is universal and determined independent of the condensate.
Here we look into the phase transition from the side of the chiral-restored phase.
Since the order parameter consists of the scalar or pseudoscalar condensate,
the effective potential can be written by such a condensate in a symmetric form in the chiral-restored phase. 
The $R$-boundary may be then found by the analysis of this effective potential,
while we cannot see what types of condensates will be realized after the phase transition.
In the SSB phase, on the other hand, the effective potential may be written in terms of only the scalar condensate,
so that the $L$-boundary can have different predictions.
Within the MFA, it has been shown that the chiral-restored phase undergoes the second-order phase transition at the $R$-boundary.
We study the nature of the inhomogeneous phase transition around the $R$-boundary by looking into both quantum and thermal fluctuations of quark-antiquark pairs or quark particle-hole pairs (hereinafter collectively called ``chiral pairs'') in the chiral-restored phase.

A similar situation also arises in the context of pion condensation in nuclear matter, where nucleon particle-hole pairs are excited~\cite{mig}.
In condensed matter physics, it corresponds to the FFLO state in superconductivity, where electron Cooper pairs are excited~\cite{super}.
One important common feature prevailing in these phenomena can be seen through the dispersion relation of the fluctuations which has a minimum at a nonzero momentum $|\bm{q}|=q_c$ on the two-dimensional sphere\footnote{
In this case, the fluctuations become soft on a finite manifold in momentum space, rather than at a single point~\cite{bra}.}
in isotropic systems~\cite{bra,dyu,hoh,fre,oha}, which suggests that the order parameter is spatially modulated after the phase transition.
This is qualitatively different from the usual phase transitions, such as homogeneous transitions in superconductivity or those for chiral symmetry breaking, where the spectrum of the fluctuations has a nonzero minimum at $|\bm{q}|=0$.
The effect of the fluctuations in the vicinity of the critical point has been studied by Nozi{\`e}re and Schmitt-Rink~\cite{nsr} within the linear (Gaussian) approximation to clarify the BCS-BEC crossover problem.
It has been further discussed in the context of a BEC of atoms~\cite{gri}, and also studied to understand a precursor of (color-)superconductivity, known as the pseudogap phenomenon \cite{lok,kit}.

A general theory for the inhomogeneous phase transition has been first presented by Brazovskii~\cite{bra} at finite temperature.
A similar issue has been discussed by Dyugaev~\cite{dyu} at zero temperature in the context of pion condensation.
They have taken into account the interactions among the fluctuations beyond the Gaussian approximation.
Unlike the homogeneous phase transition, such a nonlinear effect is now essential.
One of the remarkable findings is the change of the order of the phase transition stemming from the fluctuation effects;
the second-order phase transition within the MFA is changed to the first-order one (sometimes termed the {\it fluctuation-induced first-order phase transition}).
This subject has been further studied within the renormalization group approach~\cite{lin}.
Also, the Brazovskii theory has been applied to diblock coplymers~\cite{lei,fre,hoh}, including its experimental verification~\cite{bat}.
However, it seems that the importance of such studies is not fully conceded, e.g., in the discussion of the FFLO state.
In the previous work~\cite{kar}, two of the authors (T.T. and T.-G.L.) have presented a heuristic argument about the fluctuation-induced first-order phase transition for the inhomogeneous chiral transition, which we called the {\it Brazovskii-Dyugaev effect}.
In this paper, extending this work to the general case with $O(N)$ symmetry, we elucidate the particular roles of quantum and thermal fluctuations.
We also point out a continuity of the effects of both fluctuations across the phase boundary, by analyzing the behavior of the correlation function attributed to the excitations of the Nambu-Goldstone modes in iCP.

Another purpose of the present paper is to draw one's attention again to the fluctuation-induced first-order phase transition.
Throughout the paper we emphasize some common features for inhomogeneous phase transitions, such as those into the FFLO state in superconductivity. 
We also discuss some observational implications peculiar to the fluctuation-induced first-order phase transition.
Recently, in a B20 compound MnSi which undergoes a fluctuation-induced first-order transition of the Brazovsii type\footnote{
This type is relevant for the case with an $O(N)$ symmetric $N$-component order parameter, which differs from the case with an order parameter coupled to a fluctuating gauge field (e.g., for superconductors and smectic-A liquid crystals~\cite{hal}) or with sufficiently large components $N \geq 4$~\cite{bak} (see also, e.g., Ref.~\cite{bin}).}
\cite{bra}, an unequivocal experimental confirmation has been obtained via neutron scattering and thermodynamic observables~\cite{jan}. 
The first-order character of such a transition may also be expected to be experimentally confirmed for the inhomogeneous chiral transition, e.g., in relativistic heavy-ion collisions.

The paper is organized as follows.
In Sec.\,\ref{sec:framework}, we give a theoretical framework for analyzing the effect of fluctuations around the order parameter in the inhomogeneous chiral transition.
After that, in Secs.,\ref{sec:nonlinear} and \ref{sec:BD}, we discuss the nonlinear effects of fluctuations and the Brazovskii-Dyugaev effect, respectively, before the argument of anomalies in the thermodynamic quantities in Sec.\,\ref{sec:anomalies}.
Finally, Sec.\,\ref{sec:summary} concludes with a summary and remarks.

\section{Framework}
\label{sec:framework}

For the following Lagrangian density in the two-flavor Nambu--Jona-Lasinio (NJL) model in the chiral limit, with a quark field for two flavors $\psi$, Pauli matrices in isospin space $\btau$, and a coupling constant $G$,
\beq
{\cal L} = {\bar\psi}i\ds{\partial}\psi + G\left[\left({\bar\psi}\psi\right)^2 + \left({\bar\psi}i\gamma_5\btau\psi\right)^2\right],
\eeq
 the partition function reads $Z=\int{\cal D}\psi\int{\cal D}{\bar\psi}e^{-S}$
 with the Euclidean action in imaginary time ($t\rightarrow -i\tau$) being
\beq
S = -\int_0^\beta d\tau\int d^3\bm{x} \left[
 {\bar\psi}\left(-\gamma^0\frac{\partial}{\partial\tau}+i\bgamma\cdot\nabla+\mu\gamma^0\right)\psi
 + G\left[\left({\bar\psi}\psi\right)^2+\left({\bar\psi}i\gamma_5\btau\psi\right)^2\right] \right],
\eeq
 where $\beta=1/T$ is the inverse temperature and $\mu$ is the chemical potential.
Introducing the auxiliary fields
 $\phi_a=(-2G{\bar\psi}\psi, -2G{\bar\psi}i\gamma_5\btau\psi)$,
 one can write the Euclidean partition function as
\beqa
Z &=&
 \int{\cal D}\psi\int{\cal D}{\bar\psi}\int\prod_a{\cal D}\phi_a
 \exp\left[\int_0^\beta d\tau \int d^3\bm{x}
  \left[{\bar\psi}\left(-\gamma^0\frac{\partial}{\partial\tau}
  + i\bgamma\cdot\nabla+\mu\gamma^0-(\phi_0+i\gamma_5\btau\cdot\bphi)\right)\psi
  -\frac{1}{4G}\phi_a^2\right]\right] \nonumber \\
&=& \int\prod_a{\cal D}\phi_a e^{-S_0},
\eeqa
 where the effective action is
\beq
S_0 = \int_0^\beta d\tau\int d^3\bm{x} \left[\frac{1}{4G}\phi_a^2\right]
 - {\rm Tr}{\rm log}\left[-G_F^{-1}\right],
\eeq
 with
\beqa
G_F^{-1}
&=& - \gamma^0\frac{\partial}{\partial\tau}+i\bgamma\cdot\nabla+\mu\gamma^0
 - \left(\phi_0+i\gamma_5\btau\cdot\bphi\right) \nonumber \\
&\equiv& S_\beta^{-1}-\Delta , 
\eeqa
 and $\Delta\equiv(\phi_0+i\gamma_5\btau\cdot\bphi)$.
The inverse of the thermal Green's function, $S_\beta^{-1}$,
 can be written as $S_\beta^{-1}(i\nu_m,\bm{p})=\ds{p}$ with $p_0=i\nu_m+\mu$
 in the frequency and momentum representation,
 where $\nu_m=(2m+1)\pi T$ is the Matsubara frequency for fermions.
Thus we find
\beqa
S_0(\phi)
&=& \int_0^\beta d\tau\int d^3\bm{x}\left[\frac{1}{4G}\phi_a^2\right]
 - {\rm Trlog}[S^{-1}_\beta] - \frac{1}{2}{\rm Tr}[\Delta S_\beta]^2
 - \frac{1}{4}{\rm Tr}[\Delta S_\beta]^4 + \cdots \nonumber \\ 
&=& S_f + \frac{1}{2}T\sum_{n_1}\int\frac{d^3\bm{q}_1}{(2\pi)^3}
 \Gamma_{\rm ps}^{(2)}(i\omega_{n_1},\bm{q}_1)
 \phi_a(i\omega_{n_1},\bm{q}_1)\phi_a(-i\omega_{n_1},-\bm{q}_1) \nonumber \\
&&
 + \frac{1}{4!}T^4 \prod_{i=1}^4\sum_{n_i}\int\frac{d^3\bm{q}_i}{(2\pi)^3}
 {\hat\lambda}(\{i\omega_{n_i}\},\{\bm{q}_i\})
 \phi_a(i\omega_{n_1},\bm{q}_1)\phi_a(i\omega_{n_2},\bm{q}_2)
 \phi_b(i\omega_{n_3},\bm{q}_3)\phi_b(i\omega_{n_4},\bm{q}_4) + \cdots, 
\label{effact}
\eeqa
 where $\omega_n=2\pi nT$ is the Matsubara frequency for bosons
 and $S_f$ is the action for free quarks.
Since the component $(\omega_n=0$, $|\bm{q}|=q_c)$
 is the relevant degree of freedom,
 we here approximate the vertex function $\hat\lambda$
 by the local four-point function
 ${\hat\lambda}(\{i\omega_{n_i}\},\{\bm{q}_i\}) = \lambda(2\pi)^3
 \delta(\bm{q}_1+\bm{q}_2+\bm{q}_3+\bm{q}_4)
 \delta(\omega_{n_1}+\omega_{n_2}+\omega_{n_3}+\omega_{n_4})$
 with a coupling constant $\lambda$.
As we shall see later,
 we must keep the frequency dependence of the composite fields $\phi_a$
 to extract the correct behavior of the thermodynamic quantities at $T=0$. 
The above effective action is obviously
 $SU(2)\times SU(2)\simeq O(4)$ symmetric in the chiral-restored phase.
The $q\bar q$ polarization function $\Pi^0_{\rm ps}(i\omega_n,\bm{q})$
 and the inverse two-point function $\Gamma_{\rm ps}^{(2)}(i\omega_{n},\bm{q})$
 are defined, respectively, by 
\beqa
&& \Pi^0_{\rm ps}(i\omega_n,\bm{q}) = - N_fN_cT\sum_m\int\frac{d^3 \bm{p}}{(2\pi)^3}
 {\rm tr}\left[i\gamma_5\tau_3S_\beta(i\omega_n+i\nu_m,\bm{q}+\bm{p})
 i\gamma_5\tau_3S_\beta(i\nu_m,\bm{p})\right],
 \label{pol} \\
\mbox{and}
&& \Gamma_{\rm ps}^{(2)}(i\omega_{n},\bm{q})
 = \frac{1-2G\Pi^0_{\rm ps}(i\omega_n,\bm{q})}{2G}.
 \label{tpf}
\eeqa
Within the linear approximation for the fluctuations,
 only the first two terms are sufficient in Eq.~(\ref{effact}),
 without further terms which give the nonlinear effects
 coming from the mutual interactions of fluctuations,
 such as the fourth-order term. 
In the following discussions, however,
 we keep the terms up to fourth order in $\phi_a$, 
 as in a model {\it \`{a} la} Brazovskii~\cite{bra,hoh,fre}.

\section{Nonlinear effects of fluctuations}
\label{sec:nonlinear}

We first consider the thermodynamic potential $\Omega=-T{\rm log}Z$
 within the linear approximation:
\beq
\Omega_{\rm LA}
 = \Omega_f + 2TV\sum_n\int\frac{d^3\bm{q}}{(2\pi)^3}
 \ln\left[1-2G\Pi_{\rm ps}^0(i\omega_n,\bm{q})\right],
 \label{lin}
\eeq
 where $\Omega_f$ is the thermodynamic potential for free quarks
 and $V$ is the volume of the system.
This thermodynamic potential corresponds to that obtained by
 Nozi\`eres and Schmitt-Rink~\cite{nsr} for superconductivity.
Unlike the homogeneous phase transition,
 the polarization function has a minimum at $|\bm{q}|=q_c\neq 0$,
 i.e., $\frac{\partial\Pi^0_{\rm ps}(0,\bm{q})}{\partial |\bm{q}|}|_{|\bm{q}|=q_c}=0$,
 for the case of the inhomogeneous transition in isotropic systems.
Correspondingly, the criterion {\it \`{a} la} Thouless~\cite{tho},
 $1-2G\Pi_{\rm ps}^0(i\omega_n=0,q_c)=0$, can be derived
 as the threshold condition within the MFA.
This condition is equivalent to vanishing
 of the coefficient of the second-order term in Eq.~(\ref{effact}).

Next we shall see that 
 the nonlinear effects become essential for the inhomogeneous phase transition,
 which differs from the usual phase transition.
Since the effective action is chiral symmetric,
 we can choose the thermal average of the pseudoscalar field
 to be $\beta\Phi(\bm{q})\delta_{n0}=\langle\phi_3\rangle$,
 as an appropriate order parameter for the inhomogeneous chiral transition.
Then the thermodynamic potential can be expressed in powers of $\Phi$,
 after putting $\phi_a=\langle\phi_a\rangle+\xi_a$
 and integrating out the fluctuation fields $\xi_a$,
\beqa
\Omega
&=& \Omega_0+\frac{1}{2!}
\prod_{i=1}^2 \int\frac{d^3\bm{q}_i}{(2\pi)^3}{\bar \Gamma}_{\rm ps}^{(2)}(\{\bm{q}_i\})\Phi(\bm{q}_1)\Phi(\bm{q}_2) \nonumber \\
&&
 + \frac{1}{4!} 
\prod_{i=1}^4 \int\frac{d^3\bm{q}_i}{(2\pi)^3}{\bar \Gamma}_{\rm ps}^{(4)}(\{\bm{q}_i\})\Phi(\bm{q}_1)\Phi(\bm{q}_2)\Phi(\bm{q}_3)\Phi(\bm{q}_4) \nonumber \\
&&
 + \frac{1}{6!} 
\prod_{i=1}^6 \int\frac{d^3\bm{q}_i}{(2\pi)^3}{\bar \Gamma}_{\rm ps}^{(6)}
 (\{\bm{q}_i\})\Phi(\bm{q}_1)\Phi(\bm{q}_2)\Phi(\bm{q}_3)
 \Phi(\bm{q}_4)\Phi(\bm{q}_5)\Phi(\bm{q}_6) + \cdots, \label{thpot}
\eeqa
 where each coefficient includes the effects of fluctuations.
The first term represents the ring diagrams (bubbles),
 while the quantities of $\Pi_{\rm ps}^0$
 are modified by the fluctuations, as we will see below.

\subsection{Propagator of a chiral pair fluctuation field}
\label{subsec:propagator}

By using the polarization function,
 we can construct the propagator within the random-phase approximation (RPA).
The polarization function defined in Eq.~(\ref{pol})
 can be written in an apparent form~\cite{abr,fet},
 with the Fermi-Dirac distribution function $f(\epsilon)=(1+e^{\beta\varepsilon})^{-1}$,
\beqa
\Pi_{\rm ps}^0(i\omega_n,\bm{q})
&=& N_fN_c\sum_{\bm{p}} \left[(f(|\bm{p}|+\mu)-f(|\bm{p}+\bm{q}| +\mu))
 \frac{1-\bm{p}\cdot(\bm{p}+\bm{q})/|\bm{p}||\bm{p} 
 +\bm{q}|}{i\omega_n+|\bm{p}+\bm{q}|-|\bm{p}|}\right. \nonumber\\
&& +(f(|\bm{p}+\bm{q}|-\mu)-1+f(|\bm{p}|+\mu))
 \frac{1+\bm{p}\cdot(\bm{p}+\bm{q})/|\bm{p}||\bm{p}+\bm{q}|}
 {i\omega_n-|\bm{p}+\bm{q}|-|\bm{p}|} \nonumber\\
&& +(1-f(|\bm{p}+\bm{q}|+\mu)-f(|\bm{p}|-\mu))
 \frac{1+\bm{p}\cdot(\bm{p}+\bm{q})/|\bm{p}||\bm{p}+\bm{q}|}
 {i\omega_n+|\bm{p}+\bm{q}|+|\bm{p}|}\nonumber\\
&& \left.(f(|\bm{p}+\bm{q}|-\mu)-f(|\bm{p}|-\mu))
 \frac{1-\bm{p}\cdot(\bm{p}+\bm{q})/|\bm{p}||\bm{p}+\bm{q}|}
 {i\omega_n-|\bm{p}+\bm{q}|+|\bm{p}|}\right] ,
\label{polf}
\eeqa
which consists of the vacuum contribution,
 $\Pi_{\rm ps}^0(i\omega_n,\bm{q})|_{(\mu,T) \rightarrow 0}$,
 and the remaining medium contribution.
Here the ultraviolet divergence of the vacuum contribution
 should be regularized by the proper time regularization (PTR),
 whose explicit form is described in Ref.~\cite{kar}.
Each term in Eq.~(\ref{polf}) may be easily understood
 in terms of particle-antiparticle and particle-hole excitations, 
 where the last term corresponds to the thermal Lindhard function
 within the nonrelativistic approximation
 (for details of the derivation, see Appendix~\ref{sec:a}).
Note here that the following properties hold:
 $\Pi_{\rm ps}^0(i\omega_n,\bm{q})=\Pi_{\rm ps}^0(-i\omega_n,\bm{q})$
 and $\Pi_{\rm ps}^0(i\omega_n,\bm{q})=\Pi_{\rm ps}^0(i\omega_n,-\bm{q})$.

By the proper analytic continuation
 $\Pi_{\rm ps}^0(i\omega_n\rightarrow \omega+i\eta,\bm{q})$,
 the polarization function can be written as
\begin{align}
\Pi_{\rm ps}^0(i\omega_n,\bm{q})
 = {\rm Re}\Pi_{\rm ps}^{0}(\omega+i\eta,\bm{q})|_{\omega=i\omega_n}
 + i\, {\rm sign}(\omega_n) {\rm Im}\Pi_{\rm ps}^{0}(\omega+i\eta,\bm{q})|_{\omega=i\omega_n}.
\end{align}
Here we 
 evaluate the imaginary part (for details we refer the reader to Appendix~\ref{sec:b}):
\begin{align}
{\rm Im}\Pi_{\rm ps}^0(\omega+i\eta,\bm{q})
 &= N_fN_cT\frac{\omega^2 - q^2}{8\pi q} \Big\{
 \ln\left[ 1 + e^{-\beta(q/2 + \mu + \omega/2)} \right]
 - \ln\left[ 1 + e^{-\beta(q/2 + \mu - \omega/2)} \right] \notag \\
 &~~~~~~~~~~~~~~~~~~~
 + \ln\left[ 1 + e^{-\beta(q/2 - \mu + \omega/2)} \right]
 - \ln\left[ 1 + e^{-\beta(q/2 - \mu - \omega/2)} \right]
 + \beta\omega\theta(|\omega|-q) \Big\} ,
\label{ipol}
\end{align}
 while we have numerically found it
 in the previous paper~\cite{kar}\footnote{We have kept
 the two terms in the order of $\omega$ and $\omega^2$ in the previous work,
 but the term with $\omega^2$ is not necessarily needed.}.

The Green's function of chiral pair fluctuation fields,
 $G_{\rm ps}(i\omega_n,\bm{q})$,
 can be then defined by the use of the two-point function~(\ref{tpf}):  
\beq
 G_{\rm ps}(i\omega_n,\bm{q})
\equiv g_{\phi qq}^{-2}[\Gamma_{\rm ps}^{(2)}(i\omega_{n},\bm{q})]^{-1},
\eeq
 where $g_{\phi qq}$ is an effective coupling constant
 between quarks and a fluctuation field~\cite{kle}.
Since the behavior around $|\bm{q}|=q_c$ and $\omega_n=0$
 is important in the vicinity of the phase boundary,
 we expand it as
\beq
 G_{\rm ps}^{-1}(i\omega_n,\bm{q})
 \sim \tau+\gamma\left(|\bm{q}|^2-q_c^2\right)^2+\alpha|\omega_n| ,
\label{gfn}
\eeq
 where $\tau=G^{-1}_{\rm ps}(0,|\bm{q}|=q_c)$,
 $\gamma=\frac{1}{2}d^2G^{-1}_{\rm ps}(0,|\bm{q}|=q_c)/(d|\bm{q}|^2)^2$,
 and $\alpha=g_{\phi qq}^2d{\rm Im}\Pi^0_{\rm ps}(\omega=0,|\bm{q}|=q_c)/d\omega$.

\subsection{Thermodynamic potential}
\label{sec:thermo}

Using the effective action with the background field method,
 we can evaluate the fluctuation effects\footnote{We evaluate
 the thermodynamic potential in an $O(4)$ symmetric way,
 while we have discarded other $\phi_a$ (except for $\phi_3$)
 in the previous paper~\cite{kar}.}.
Inserting $\phi_a=\beta\Phi(\bm{q})\delta_{n0}\delta_{a3}+\xi_a$ into Eq.~(\ref{effact}),
 the effective action can be written as
\beq
 S_0(\phi_a) = S_0(\Phi)+S_1(\Phi,\xi_a).
\eeq
Accordingly,
 the thermodynamic potential is given by the functional integral:
\beq
 \Omega (\Phi) = TS_0(\Phi)-T{\rm log}\int\prod_a{\cal D}\xi_a{\rm exp}[-S_1(\Phi,\xi_a)].
\eeq

Each vertex function in Eq.~(\ref{thpot}) is then given by
\beq
{\bar \Gamma}^{(n)}(\bm{q}_1,\bm{q}_2, \cdots, \bm{q}_n)
 = (2\pi)^{3n}\left.\frac{\delta^n\Omega}
 {\delta\Phi(-\bm{q}_1)\delta\Phi(-\bm{q}_2)\cdots\delta\Phi(-\bm{q}_n)}\right|_{\Phi=0} .
\eeq
The key equation is the first functional derivative,
\begin{align}
(2\pi)^3\frac{\delta\Omega}{\delta\Phi\left(-\bm{q}_1\right)}
=&~G_{\rm ps}^{-1} \left(0,\bm{q}_1\right) \Phi\left(\bm{q}_1\right)
 + \frac{\lambda}{3!} \int \frac{d^3 \bm{q}_2~ d^3 \bm{q}_3}{(2\pi)^6}
 \Phi\left(\bm{q}_2\right)\Phi\left(\bm{q}_3\right)
 \Phi\left(\bm{q}_1-\bm{q}_2-\bm{q}_3\right) \notag \\
 & + \frac{\lambda}{2!}T^2\sum_{n}\int\frac{d^3 \bm{q}_2 ~d^3 \bm{q}_3}{(2\pi)^6}
 \left[\langle \xi_3\left( i\omega_n,\bm{q}_2 \right)\xi_3
 \left(-i\omega_n,\bm{q}_3 \right)\rangle_{\xi}
 + \langle \xi_0\left( i\omega_n,\bm{q}_2 \right)\xi_0
 \left(-i\omega_n,\bm{q}_3 \right)\rangle_{\xi}\right]
 \Phi\left(\bm{q}_1-\bm{q}_2-\bm{q}_3\right),
\label{key}
\end{align}
where the symbol $\langle\cdots\rangle_\xi$ denotes the thermal average,
 and we have used the following relation:
$\langle\xi_0\left( i\omega_n,\bm{q}_2 \right)\xi_0\left( -i\omega_n,\bm{q}_3 \right)\rangle_{\xi}
=\langle\xi_1\left( i\omega_n,\bm{q}_2 \right)\xi_1\left( -i\omega_n,\bm{q}_3 \right)\rangle_{\xi}
=\langle\xi_2\left( i\omega_n,\bm{q}_2 \right)\xi_2\left( -i\omega_n,\bm{q}_3 \right)\rangle_{\xi}$.
In general, the thermal average
 $\langle \xi (i\omega_n,\bm{q}_2) \xi (-i\omega_n, \bm{q}_3) \rangle_\xi$
 has the off-diagonal momentum components,
 but we can neglect such components
 as long as the loop integrals are concerned~\cite{bra}.
Thus,
\beq
\langle\xi_a (i\omega_n,\bm{q}_2) \xi_a (-i\omega_n, \bm{q}_3)\rangle_\xi
 = \beta(2\pi)^3\delta(\bm{q}_2+\bm{q}_3){\bar G}_{a}(i\omega_n,\bm{q}_2).
\eeq
where $\bar G_{a}(i\omega_n,\bm{q})$ is the self-consistent Green's function, given by
 ${\bar G}_{a}(i\omega_n,\bm{q})=[r_a+\gamma(|\bm{q}|^2-q_c^2)^2+\alpha|\omega_n|]^{-1}$
 with
\beqa
r_3 &=&
 \tau + V^{-1} \frac{\lambda}{2} \int \frac{d^3 \bm{q}}{(2\pi)^3} \Phi \left( \bm{q} \right)\Phi \left( -\bm{q} \right) +   \frac{\lambda}{2} T \sum_{n} \int \frac{d^3 \bm{q}}{(2\pi)^3}
 \left[{\bar G_3(i\omega_n,\bm{q})}+{\bar G_0(i\omega_n,\bm{q})}\right]\nonumber\\
&\equiv&\tau + V^{-1} \frac{\lambda}{2} \int \frac{d^3 \bm{q}}{(2\pi)^3} \Phi \left( \bm{q} \right)\Phi \left( -\bm{q} \right) +   \frac{\lambda}{2}\left(I_1(r_3)+I_1(r_0)\right).
\label{SD}
\eeqa
The integrals $I_n$ are defined by
\beq
I_n(r) = T\sum_k \int\frac{d^3\bm{q}}{(2\pi)^3}
 \left(\frac{\partial^{(n-1)}}{\partial r^{n-1}}\right)
 \frac{1}{r+\gamma(|\bm{q}|^2-q_c^2)^2+\alpha|\omega_k|},
\eeq
where the integrals with $n\leq2$ should be regularized
 by some regularization methods (see Appendix~\ref{sec:c} for details).
Similarly, $r_0(=r_1=r_2)$ reads
\beq
 r_0 = \tau + V^{-1}\frac{\lambda}{6}\int\frac{d^3 \bm{q}}{(2\pi)^3}
 \Phi\left( \bm{q} \right)\Phi\left( -\bm{q} \right)
 + \frac{\lambda}{2}\left(I_1(r_3)+I_1(r_0)\right).
\label{SD0}
\eeq
While, strictly speaking, there are other two diagrams
 contributing to $r_3$ and $r_0$,
 their contribution can be neglected
 in the region $r_{3,0}^{1/2}\ll q_c$~\cite{bra}.
Using $r_3$,
 Eq.~(\ref{key}) can be rewritten as 
\begin{align}
(2\pi)^3\frac{\delta\Omega}{\delta\Phi\left(-\bm{q}_1\right)}
=& \left[ r_3 + \gamma \left( |\bm{q}_1|^2 - q_c^2 \right)^2 \right]\Phi\left(\bm{q}_1\right)
 -V^{-1} \frac{\lambda}{2!} \Phi \left( \bm{q}_1 \right)
 \int\frac{d^3 \bm{q}_2}{(2\pi)^3}\Phi\left(\bm{q}_2\right)\Phi\left( -\bm{q}_2\right) \notag \\
&+ \frac{\lambda}{3!}\int\frac{d^3 \bm{q}_2~ d^3 \bm{q}_3}{(2\pi)^6}\Phi\left(\bm{q}_2\right)
 \Phi\left(\bm{q}_3 \right)\Phi\left(\bm{q}_1-\bm{q}_2-\bm{q}_3\right). 
\label{key2}
\end{align}
Thus, $\Gamma^{(1)}$ is obviously vanished as should be expected.
Subsequent derivatives of $\Omega$ give the even-order vertex functions.
Note here that $r_a$ is a functional of $\Phi$ 
 and their derivatives satisfy the following equations:
\beqa
\frac{\delta r_3}{\delta\Phi(-\bm{q}_2)}
&=& \frac{(2\pi)^{-3}V^{-1}\lambda\Phi(\bm{q}_2)}{1-\frac{\lambda}{2}I_2(r_3)}
 +\frac{\frac{\lambda}{2}I_2(r_0)}{1-\frac{\lambda}{2}I_2(r_3)}
 \cdot\frac{\delta r_0}{\delta\Phi(-\bm{q}_2)}, \nonumber\\
\text{and}\quad
\frac{\delta r_0}{\delta\Phi(-\bm{q}_2)}
&=& \frac{(2\pi)^{-3}V^{-1}\frac{\lambda}{3}\Phi(\bm{q}_2)}{1-\frac{\lambda}{2}I_2(r_0)}
 +\frac{\frac{\lambda}{2}I_2(r_3)}{1-\frac{\lambda}{2}I_2(r_0)}
 \cdot\frac{\delta r_3}{\delta\Phi(-\bm{q}_2)}.
\eeqa
The second-order vertex function thus reads
\beq
{\bar \Gamma}^{(2)}(\bm{q}_1,\bm{q}_2)
 = (2\pi)^3\delta(\bm{q}_1+\bm{q}_2)(\tau_R+\gamma(|\bm{q}|^2-q_c^2)^2),
\label{gam2}
\eeq
where $\tau_R=r_3(\Phi=0)=r_0(\Phi=0)$.
Likewise, the fourth-order vertex function is
\beq
{\bar\Gamma}^{(4)}(\{\bm{q}_i\})
= (2\pi)^3\lambda \left[\delta(\bm{q}_1+\bm{q}_2+\bm{q}_3+\bm{q}_4)
 +(2\pi)^3V^{-1}\frac{\frac{2\lambda}{3}I_2(\tau_R)}{1-\lambda I_2(\tau_R)}
 \left[\delta(\bm{q}_1+\bm{q}_2)\delta(\bm{q}_3+\bm{q}_4)
 +2~{\rm permutations}\right]\right].
\label{g4}
\eeq

\section{Brazovskii-Dyugaev effect}
\label{sec:BD}

\subsection{Fluctuation-induced first-order phase transition}
\begin{figure*}
\centering
\includegraphics[width=7cm]{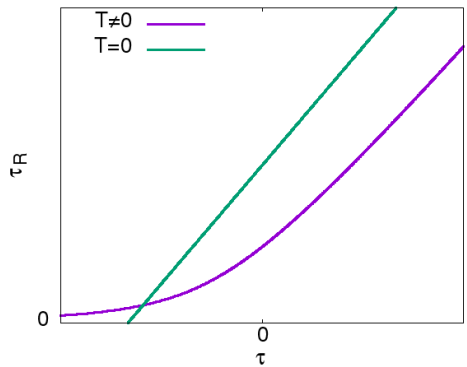}
\caption{The $\tau$ dependences of $\tau_R$
 at $T\neq0$ (purple curve) and $T=0$ (green line).
 $\tau_R$ always takes positive values when $T\neq0$,
 while, when $T=0$, it vanishes
 at $\tau=-\frac{\lambda\Lambda^3}{48\alpha\pi^{5/2}\gamma^{3/2}q_c^3}$.}
\label{tau}
~\\
\end{figure*}

First, we consider the second-order term (\ref{gam2}).
If $\tau_R$ becomes zero,
 it should be a signal of the second-order phase transition.
From Eq.~(\ref{SD}), $\tau_R$ satisfies 
\beqa
\tau_R &=& \tau + \lambda I_1(\tau_R), \nonumber\\
\text{with}\quad
\tau~ &=& \tau_R - \frac{\lambda T}{2\pi^2}\int_{\tau_R/\Lambda^2}^\infty{ds}
 \left[\frac{\tau_R^{1/2}}{2}\left(\frac{\pi}{(4\gamma q_c^2s)^3}\right)^{1/2}
 +\tau_R^{-1/2}\left(\frac{\pi q_c^2}{4\gamma s}\right)\right]
 e^{-s}{\rm coth}\left(\frac{\alpha\pi Ts}{\tau_R}\right).
\eeqa
Looking into the behavior around $\tau_R=0$, we find that
\beqa
\tau &\simeq& \tau_R - \frac{\lambda T q_c}{4\pi\gamma^{1/2}\tau_R^{1/2}}
~~~{\rm for}~~T\neq 0\label{taut}, \\
\text{while}\quad
\tau &\simeq& \tau_R-\frac{\lambda\Lambda^3}{48\alpha\pi^{5/2}\gamma^{3/2}q_c^3}
~~~{\rm for}~~T=0 .
\label{tau0}
\eeqa
This is due to the singularity of $G_{\rm ps}^R(i\omega_n,\bm{q})$
 on the sphere $|\bm{q}|=q_c$,
 which is a common feature for inhomogeneous phase transitions
 in isotropic systems~\cite{bra,dyu,hoh,fre,oha}.
From Eq.~(\ref{taut}) and Fig.~\ref{tau}, we can see that
 $\tau$ diverges at $\tau_R=0$
and $\tau_R$ is always positive for all range of $\tau$, which implies that
the phase transition is prohibited at finite temperature.
On the other hand, there is no divergence at zero temperature.
In addition, the point $\tau_R=0$ at $T=0$ is somewhat shifted fromthe point $\tau=0$.
The difference of  $T\neq0$ and $T=0$
 can be easily understood from the fact that
 the lowest Matsubara frequency is dominant
 and the leading behavior (\ref{taut}) can be obtained
 by putting $\omega_n=0$ into the integral $I_2$.
Thus we can observe that
 there takes place a kind of dimensional reduction 
 from $1+3$ to $0+3$ dimensions at $T\neq 0$.
It would be interesting to see a similarity
 to the Coleman-Mermin-Wagner theorem~\cite{cmw},
 which claims that the lower critical dimension is $1+2$
 for thermal fluctuations~\cite{wen}.
In the case of $T=0$,
 the imaginary part in $\bar G_{\rm ps}$ becomes important
 to lead to no divergent behavior; 
 quantum fluctuations are gentle
 and only shifts the critical point\footnote{The logarithmic divergence at $T=0$
 was claimed by Kleinert \cite{klei}, but it should be remedied by the proper treatment of the imaginary part included in the Green function.}.

The above considerations are insufficient for the possibility of the phase transition
 due to the only consideration for the second-order phase transition.
Next, we shall introduce the fourth-order and sixth-order vertex functions
 to see whether the system undergoes the first-order phase transition.
The sign change of the fourth-order vertex function by fluctuations
 has been first shown by Brazovskii~ at $T\neq 0$ and Dyugaev at $T=0$. 
The integral $I_2(\tau_R)$ in $\bar\Gamma^{(4)}$ which will be seen in Eq.~(\ref{gam4}) can be evaluated as
\beq
I_2(\tau_R) = -\frac{T}{2\pi^2}\int_0^\infty ds\left[\frac{1}{2}\left(\frac{\pi}{(4\gamma q_c^2)^3s}\right)^{1/2}\tau_R^{-1/2}+\left(\frac{\pi q_c^2s}{4\gamma}\right)^{1/2}\tau_R^{-3/2}\right]e^{-s}{\rm coth}\left(\frac{\alpha\pi Ts}{\tau_R}\right).
\eeq
Looking into the behavior around $\tau_R=0$, we find that
\beqa
 I_2(\tau_R) &\simeq& -\frac{Tq_c}{8\pi\gamma^{1/2}}\tau_R^{-3/2}
~~~{\rm for}~~T\neq 0 \label{i2t}, \\
\text{while}\quad
I_2(\tau_R) &\simeq& -\frac{q_c}{4\alpha\pi^2\gamma^{1/2}}\tau_R^{-1/2}
~~~{\rm for}~~T=0.
\label{i20}
\eeqa
This result shows that
 the effects of fluctuations lead to the divergence of the integral $I_2$
 near the phase boundary.
Unlike $\tau_R$,
 quantum fluctuations also give rise to a singular behavior,
 while it is less drastic than thermal fluctuations.
These features can be understood as for the case of $\tau_R$.
The expression (\ref{g4}) is physically given
 by summing up the ``dangerous diagrams,''
 which are composed of bubbles
 of the renormalized propagator in the chiral-restored phase,
\beqa
G_{\rm ps}^R(i\omega_n,\bm{q})
&=& {\bar G}_{\rm ps}(i\omega_n,\bm{q};\tau=\tau_R) \nonumber\\
&=& \frac{1}{\tau_R+\gamma(|\bm{q}|^2-q_c^2)^2+\alpha|\omega_n|} ,
\label{gr}
\eeqa
 and also represent the long-range interaction among chiral pair fluctuations,
\beq
L(\bm{k}) = T\sum_n\int\frac{d^3\bm{q}}{(2\pi)^3}
 G_{\rm ps}^R(i\omega_n,\bm{q})G_{\rm ps}^R(-i\omega_n,\bm{k}-\bm{q}).
\eeq
We can easily see that
 $L(\bm{k})$ becomes the most singular
 and $L(\bm{k})\rightarrow - I_2(\tau_R)$ as $k\rightarrow 0$:
 the singularities in $G^R_{\rm ps}(i\omega_n,\bm{q})$
 and $G^R_{\rm ps}(-i\omega_n,\bm{k}-\bm{q})$
 come closer as $k\rightarrow 0$ to make the integral divergent.
Finally, we find 
\beq
{\bar\Gamma^{(4)}}
= (2\pi)^3 \lambda\frac{1+ \frac{\lambda}{3}I_2(\tau_R)}{1-\lambda I_2(\tau_R)}
 \delta(\bm{q}_1+\bm{q}_2+\bm{q}_3+\bm{q}_4), \label{gam4}
\eeq
 assuming the form of the condensate as $\Phi=\Delta{\rm sin}(q_cz)$.
Hence, the sign of ${\bar\Gamma^{(4)}}$ is changed at the point,
 $1+\frac{\lambda}{3}I_2(\tau_R)=0$,
 which suggests that the phase transition is of first order,
 i.e., the fluctuation-induced first-order phase transition.

In the above discussion,
 the $O(4)$ model including four fluctuation fields is considered.
We next expand the discussion to the $O(N)$ model for theoretical interests.
For arbitrary $N$, Eqs.~(\ref{SD}) and (\ref{SD0}) are recast into
\beqa
r_3 &=&
 \tau + V^{-1} \frac{\lambda}{2} \int \frac{d^3 \bm{q}}{(2\pi)^3}
 \Phi \left( \bm{q} \right)\Phi \left( -\bm{q} \right)
 + \frac{\lambda}{2}\left(I_1(r_3) + \frac{N-1}{3} I_1(r_0)\right), \\
r_0 &=&
 \tau + V^{-1}\frac{\lambda}{6} \int \frac{d^3 \bm{q}}{(2\pi)^3}
 \Phi \left( \bm{q} \right)\Phi \left( -\bm{q} \right)
 + \frac{\lambda}{2}\left(I_1(r_3) + \frac{N-1}{3} I_1(r_0)\right).
\eeqa
Consequently, $\bar{\Gamma}^{(4)}$ is modified as follows:
\beq
{\bar\Gamma^{(4)}}
 = (2\pi)^3 \lambda
 \frac{1+ \frac{10-N}{18}\lambda I_2(\tau_R)}{1 - \frac{N+2}{6} \lambda I_2(\tau_R)}
 \delta(\bm{q}_1+\bm{q}_2+\bm{q}_3+\bm{q}_4).
\eeq
In the case of $N=1$,
 the previous result obtained in \cite{kar} is reproduced,
 and the result for $N=4$ coincides with Eq.~(\ref{gam4}).
In the case of $N \geq 10$, on the other hand,
 the fluctuation-induced first-order phase transition does not occur
 because $\bar{\Gamma}^{(4)}$ never becomes negative.

\subsection{{\bf Continuity of the roles of fluctuations across the phase boundary}}
Here we discuss a similar feature of the fluctuations in iCP\footnote{A similar argument
 has been given for Larkin-Ovchinikov-type liquid-crystal states~\cite{rad}.}.
We have found that
 the effects of fluctuations are remarkable for the phase transition, 
 and the role of thermal fluctuations is more profound
 than that of quantum fluctuations. 
On the other hand,
 it has been shown that the fluctuations in iCP are important 
 to cause the instability of one-dimensional structures at finite temperature~\cite{lee}:
 the scalar or pseudoscalar correlation function,
 $f_a(\bm{r})=\langle \phi_a(\bm{r})\phi_a(0)\rangle$
 where $\phi_a$ denote the quark bilinear fields in scalar or pseudoscalar channel
 (i.e., $\phi_0={\bar\psi}\psi$ and $\phi_i={\bar\psi}i\gamma_5\tau_i\psi$),
 algebraically decays at large distances
 due to the low-energy Nambu-Goldstone excitations. 
It is in accord with the Landau-Peierls theorem~\cite{lan,pei}.
For definiteness, we here consider inhomogeneous chiral condensates
 of the dual chiral density wave (DCDW) type.
There appear three Naumbu-Goldstone modes ($\beta_i$)
 with the anisotropic dispersion $\omega_i^2=A_i k_z^2+B_i k_\perp^4$.
In the case of $T=0$, we find 
\beq
\langle \beta_i(\bm{r})^2\rangle_{T=0}
 = c_i\int \frac{d^4k}{(2\pi)^4}\frac{1}{k_0^2+A_ik_z^2+B_ik_\perp^4},  
\eeq
where $k_z$ is the momentum in the direction
 parallel to the wave vector of the modulation,
 while $k_\perp$ is that in the directions normal to the modulation.
Here 
 the coefficients $c_i$, $A_i$, and $B_i$ can be evaluated
 within chiral effective models~\cite{nic}.
The above integral is convergent in the infrared region.
In the case of $T\neq0$, on the other hand, we find
\beqa
 \langle \beta_i(\bm{r})^2 \rangle_{T\neq0}
&=& c_iT\sum_{\omega_n}
 \int_{-\infty}^\infty \frac{dk_z}{2\pi}\int_{l^{-1}_\perp}^{\Lambda_\perp}
 \frac{d^2\bm{k}_\perp}{(2\pi)^2}\frac{1}{\omega_n^2+A_ik_z^2+B_ik_\perp^4}, \nonumber \\
&\sim& \frac{c_iT}{8\pi \sqrt{A_iB_i}}\ln(\Lambda_\perp l_\perp) ,
\eeqa
which is divergent in the infrared region $(l_\perp^{-1} \rightarrow 0)$.
The most dominant contribution comes from the lowest Matsubara frequency $\omega_n=0$,
\beq
\langle \beta_i(\bm{r})^2 \rangle_{T\neq0}
 \sim c_iT \int_{-\infty}^\infty \frac{dk_z}{2\pi}
 \int_{l^{-1}_\perp}^{\Lambda_\perp}
 \frac{d^2\bm{k}_\perp}{(2\pi)^2} \frac{1}{A_ik_z^2+B_ik_\perp^4},
\eeq
 and exhibits an infrared singularity due to the effective {\it dimensional reduction}.
This implies that
 thermal fluctuations play a more important role
 in the infrared singularity than quantum ones.
In this way, we can see the similar features to our results obtained in the previous subsection.

The stability of the DCDW phase an also be understood in the same way.
Here the correlation function of the order parameter $\phi_3$ takes the form~\cite{lee,hid},
\beq
\langle \phi_3(z)\phi_3(0) \rangle
 \sim e^{-\frac{1}{2} \langle \left[\beta_3(z) - \beta_3(0)\right]^2 \rangle}.
\eeq
From this, we find that for $T=0$,
\beqa
\langle \left[\beta_3(z) - \beta_3(0)\right]^2 \rangle_{T=0}
&=& c_i \int \frac{d^4k}{(2\pi)^4}\frac{1-e^{ik_zz}}{k_0^2+A_ik_z^2+B_ik_\perp^4} \nonumber \\
&\sim& \frac{c_i}{16\pi\sqrt{A_iB_i}}\Lambda~~{\rm for~large}~|z|,
\eeqa
 while for $T\neq0$,
\beqa
\langle \left[\beta_3(z) - \beta_3(0)\right]^2 \rangle_{T\neq0}
&=& c_iT\sum_{\omega_n}\int \frac{d^3\bm{k}}{(2\pi)^3}
 \frac{1-e^{ik_zz}}{\omega_n^2+A_ik_z^2+B_ik_\perp^4} \nonumber \\
&\sim& \frac{c_iT}{8\pi\sqrt{A_iB_i}}
 \ln\left( A_i^{-1/2}\Lambda |z| \right)~~{\rm for~large}~|z|,
\eeqa
where the latter logarithmically diverges in the limit $|z|\rightarrow\infty$,
 and $\Lambda$ is an ultraviolet cutoff.
For details we refer the reader to Appendix~\ref{sec:d}.
Similarly, the same results can be obtained for other order parameters.
Here thermal fluctuations are still important for the same reason. 
Therefore, we can conclude that
 the correlation function algebraically decays only for $T\neq0$
 and the DCDW phase shows the feature of quasi-long range order~\cite{lee}
 (see also~\cite{hid} for RKC).
These results may suggest some continuity of the roles of fluctuations
 before and after the phase transition,
 as far as the one-dimensional modulation is concerned.

\section{Anomalies in the thermodynamic quantities}
\label{sec:anomalies}

It is well known that fluctuations affect the thermodynamic quantities
 around the phase boundary.
Anomalies in various susceptibilities (the second derivatives of the thermodynamic potential)
 are characteristic features near the critical point of the second-order phase transition;
 the specific-heat anomaly, $C_v\sim (T-T_c)^{-1/2}$, has been shown
 due to fluctuations at the critical temperature $T_c$ of superconductivity,
 while there is generated a finite discontinuity within the MFA. 
In the context of the usual chiral transition,
 the divergence of the quark number susceptibility, $\partial N/\partial\mu$,
 has been discussed~\cite{fuj}.

The singular behavior of the propagator gives rise to new types of anomalies
 in the thermodynamic quantities for the inhomogeneous phase transition.
In the context of the FFLO state in superconductivity,
 Ohashi have indicated the divergence of the electron number
 due to the fluctuation~\cite{oha},
 where $N\sim \tau^{-1/2}$ within the linear approximation,
 which means that the first derivative of the thermodynamic potential becomes singular.
As we have seen in the previous section,
 $\tau$ is renormalized to keep it to be positive definite by the non-linear effects,
 and the singularity mentioned above can not be observed.
However, its remnant should be observed.
Thus, the fluctuation-induced first-order transition is characterized
 by the discontinuity and singular behavior of the first derivative.

In the following, we shall discuss the quark number and the entropy
 for the inhomogeneous chiral transition. 
In the chiral-restored phase ($\Phi=0$),
 the thermodynamic potential is given by 
\beq
\Omega(\Phi=0)
 = \Omega_f-2TV\sum_n\int\frac{d^3\bm{q}}{(2\pi)^3}
 \ln\left(G_{\rm ps}^R(i\omega_n,\bm{q})\right),
\eeq
 which is a simple generalization of Eq.~(\ref{lin})
 with a replacement of $\tau$ by $\tau_R$.
The quark number density can be written as
\beq
 n = n_f + 2T\sum_n\int\frac{d^3\bm{q}}{(2\pi)^3}
 G_{\rm ps}^R(i\omega_n,\bm{q})
 \frac{\partial}{\partial\mu}{\bar\Pi}^0_{\rm ps}(i\omega_n,\bm{q}),
\eeq
where
\beq
n_f = \frac{N_cN_f}{\pi^2}\int_0^\infty p^2dp\left(f(p-\mu)-f(p+\mu)\right).
\eeq
Using Eq.~(\ref{gr}) and $I_1$ derived from Appendix \ref{sec:c},
{\bf the fluctuation effects then can be seen separately: 
\beqa
 n &\sim& n_f + \frac{1}{\tau_R^{1/2}} \frac{N_fN_cq_c}{8\pi^3\gamma^{1/2}}
 \int_{-\infty}^\infty pdp\frac{1}{[e^{\beta(p+\mu)/2}+e^{-\beta(p+\mu)/2}]^2}
 \left(4+\frac{q_c}{p}\ln\left|\frac{2p-q_c}{2p+q_c}\right|\right)~~~{\rm for}~~~T \neq 0, \nonumber\\
 &\sim& n_f - \frac{N_fN_c\Lambda^3}{96\pi^{9/2}\alpha\gamma^{3/2}q_c^3} \left( 4\mu + q_c \ln\left| \frac{2\mu-q_c}{2\mu+q_c}\right| \right)~~~{\rm for}~~~T=0,
\eeqa
as the leading contribution.}
Thus we can see a singular behavior at $T\neq 0$,
 while at $T=0$ only a finite gap is produced.
A similar divergence ($\propto \tau_R^{-1/2}$) is also observed in the entropy.
The entropy density is given by 
\beq
 s = s_f - 2\sum_n\int\frac{d^3\bm{q}}{(2\pi)^3}
 \ln G_{\rm ps}^R(i\omega,\bm{q})
 +2T\sum_n\int\frac{d^3\bm{q}}{(2\pi)^3}G_{\rm ps}^R(i\omega,\bm{q})
 \frac{\partial}{\partial T}{\bar\Pi}^0_{\rm ps}(i\omega_n,\bm{q}),
\eeq
where 
\beq
 s_f = \frac{2N_fN_c}{\pi^2}\int_0^\infty p dp
 \left[(2p-\mu)\ln(1+e^{-\beta(p-\mu)})+(2p+\mu)\ln(1+e^{-\beta(p-\mu)})\right].
\eeq
The second term gives a minor contribution ($\propto\ln\tau_R$),
 while the leading contribution comes from the third term,
\beq
 s \sim s_f - \frac{1}{\tau_R^{1/2}} \frac{N_fN_cq_c}{8\pi^3 \gamma^{1/2}T} \int_{-\infty}^\infty dp \frac{p(p+\mu)}{\left[ e^{\beta(p+\mu)/2} + e^{-\beta(p+\mu)/2} \right]^2} \left( 4 + \frac{q_c}{p} \ln\left| \frac{2p-q_c}{2p+q_c}\right| \right),
\eeq
 around the phase boundary.
Such a anomalous behavior may be reflected
 in the particle production during relativistic heavy-ion collisions,
 if the system crosses the phase boundary.
Here it would be worth mentioning
 that the entropy anomaly may also be a signal of the FFLO state.

\section{Summary and concluding remarks}
\label{sec:summary}

We have discussed the effects of chiral pair fluctuations
 on the inhomogeneous chiral transition by extending the previous work. 
Also, we have taken into account the non-linear effects of chiral-pair fluctuations
 in a systematic way beyond the linear approximation.
Eventually, we have elucidated
 the salient roles of quantum and thermal fluctuations separately;
 the latter is more drastic than the former
 due to the dimensional reduction,
 but both lead to the fluctuation-induced first-order phase transition.
The curvature parameter $\tau$ is renormalized
 by the fluctuation effects to be positive definite at $T\neq 0$,
 while for $T=0$ it is mildly shifted from the one within MFA.
Thus, we have observed
 that the second-order phase transition is prohibited by thermal fluctuations.
More importantly, the dangerous diagrams composed of
 the bubbles of two fluctuation Green's function become essential
 and change the sign of the fourth-order vertex function
 for both the $T=0$ and $T\neq0$ cases.
The sign of the sixth-order vertex function can be shown to be positive definite,
 and thus we can clearly see the first-order phase transition.
These features are brought about by the unique behavior
 of the dispersion of chiral pair fluctuations,
 and also common in any inhomogeneous phase transition.

It should be worth mentioning that
 the behavior of the vertex functions has been also studied
 by solving the flow equations within the renormalization group approach,
 and in addition the findings with the perturbative approach have been confirmed
 for diblock copolymers~\cite{hoh}.
The renormalization group is somewhat different from the usual treatment
 due to the existence of the special point $q_c$ in momentum space,
 but can be formulated in the similar way to the work by Shankar~\cite{sha}
 for fermion many-body systems,
 where the Fermi momentum corresponds to $q_c$.
Since our formalism is very much similar to theirs,
 one may expect that our findings are also confirmed
 by the renormalization group approach.
This is left for a future work.

The first derivative of the thermodynamic potential exhibits a singular behavior
 through the momentum integral,
 since the dispersion of chiral pair fluctuations has a minimum on the sphere $|\bm{q}|=q_c$.
To figure out such a singular behavior,
 we have evaluated the number density and entropy density,
 with the result that the fluctuation-induced first-order phase transition
 can be characterized by the discontinuity and singular behavior of the first derivatives.

Throughout the paper, we have discussed the properties of the $R$-boundary.
As for the $L$-boundary,
 it has been shown that it should be of first order in the case of DCDW,
 while of second order in the case of RKC.
Therefore, it would be interesting to apply our argument
 to the $R$-boundary of RKC,
 where the number susceptibility has been suggested
 to be divergent within the MFA~\cite{car}.

In this paper we have been concerned
 the effects of fluctuations of the order parameters.
The singular feature of the propagator of a fluctuation field
 may also affect the quark propagator~\cite{lok,kit};
 the self-energy of quarks should exhibit another anomalous behavior
 near the phase boundary.

Finally it would be worth mentioning that
 our formalism to treat the non-linear effects of fluctuations
 may also be applied to other cases,
 such as the FFLO state in superconductivity~\cite{lee2};
 the Cooper pair fluctuations are composed of the particle-particle ladder diagram instead,
 but the dispersion relation has a similar feature discussed here.
Accordingly, the entropy anomaly may be a possible evidence for the phase transition.

\section*{Acknowledgement}
We thank Y. Ohashi for stimulating discussions. 
R.Y. is supported by Grants-in-Aid for Japan Society
 for the Promotion of Science (JSPS) fellows No.~27-1814.
T.-G.L. and T.T. are partially supported by Grant-in-Aid
 for Scientific Research on Innovative Areas through No.~24105008 provided by MEXT.

\appendix

\section{Thermal Lindhard function}
\label{sec:a}
In the nonrelativistic limit,
 we consider the particle-hole polarization function (the Lindhard function at $T\neq 0$),
\beqa
\chi(q)
&=& T\sum_{\bm{p}}G\left(p+\frac{q}{2}\right)G\left(p-\frac{q}{2}\right) \nonumber\\
&=& \sum_{\bm{p}}\frac{f(\varepsilon_+)-f(\varepsilon_-)}
 {\varepsilon_+-\varepsilon_-+i\omega_n},
\label{nr-pol}
\eeqa 
where $\varepsilon_\pm=\frac{(\bm{p}\pm\bm{q}/2)^2}{2m}-\mu$.
This can be also obtained from the imaginary part of 
 the polarization function in the relativistic case
 after the analytic continuation ($i\omega_n\rightarrow \omega+i\eta$).
The imaginary part takes the form
\begin{align}
{\rm Im}\chi(\bm{q}, \omega+i\eta)
&= -\pi \sum_{\bm{p}}\left[ f(\epsilon_+) - f(\epsilon_-) \right]
 \delta(\epsilon_+ - \epsilon_- +\omega) \notag\\
&= -\frac{1}{4\pi}\int_0^\infty p^2 dp \int_{-1}^1 dx
 \left[ f\left(\zeta_p + \frac{pq}{2m}x \right) - f\left(\zeta_p - \frac{pq}{2m}x \right)\right]
 \delta\left(\frac{pq}{m}x +\omega \right) ,
\end{align}
 where $\zeta_p = \frac{p^2}{2m} + \frac{q^2}{8m} -\mu$.
Considering the argument of the delta function, we find
\begin{align}
{\rm Im}\chi(\bm{q}, \omega+i\eta)
&= -\frac{m}{4\pi q}\int_0^\infty p dp
 \left[ f\left(\zeta_p - \frac{\omega}{2} \right)
 - f\left(\zeta_p + \frac{\omega}{2} \right)\right]
 \theta\left(-\frac{pq}{m} < \omega < \frac{pq}{m} \right) \notag \\
&= -\frac{m^2T}{4\pi q}
 \left[ \ln\left\{ 1 + e^{-\beta\left(\frac{m\omega^2}{2q^2}
 + \frac{q^2}{8m} -\mu - \frac{\omega}{2} \right)} \right\}
 - \ln\left\{ 1 + e^{-\beta\left(\frac{m\omega^2}{2q^2}
 + \frac{q^2}{8m} -\mu + \frac{\omega}{2} \right)} \right\} \right].
\end{align}

\section{Imaginary part of the polarization function}
\label{sec:b}
The imaginary part of
 $\Pi_{\rm ps}^{0}(\bm{q}, \omega+i\eta)$ is given by
\begin{align}
 {\rm Im}\Pi_{\rm ps}^{0}(\bm{q}, \omega+i\eta)
 &= -N_fN_c\pi \int \frac{d^3\bm{p}}{(2\pi)^3} \bigg\{ \left[f(|\bm{p}| + \mu) - f(|\bm{p} + \bm{q}| + \mu)\right] \left[ 1 - \frac{\bm{p}\cdot(\bm{p}+\bm{q})}{|\bm{p}||\bm{p}+\bm{q}|} \right] \delta(\omega + |\bm{p} + \bm{q}| - |\bm{p}|) \notag \\
 &~~~~~~~~~~~~~~~~~~+ \left[f(|\bm{p} + \bm{q}| - \mu) - f(|\bm{p}| - \mu)\right] \left[ 1 - \frac{\bm{p}\cdot(\bm{p}+\bm{q})}{|\bm{p}||\bm{p}+\bm{q}|} \right] \delta(\omega - |\bm{p} + \bm{q}| + |\bm{p}|) \notag \\
 &~~~~~~~~~~~~~~~~~~+ \left[f(|\bm{p} + \bm{q}| - \mu) -1 + f(|\bm{p}| + \mu)\right] \left[ 1 + \frac{\bm{p}\cdot(\bm{p}+\bm{q})}{|\bm{p}||\bm{p}+\bm{q}|} \right] \delta(\omega - |\bm{p} + \bm{q}| - |\bm{p}|) \notag \\
 &~~~~~~~~~~~~~~~~~~+ \left[1 - f(|\bm{p} + \bm{q}| + \mu) - f(|\bm{p}| - \mu)\right] \left[ 1 + \frac{\bm{p}\cdot(\bm{p}+\bm{q})}{|\bm{p}||\bm{p}+\bm{q}|} \right] \delta(\omega + |\bm{p} + \bm{q}| + |\bm{p}|) \bigg\}.
\end{align}
Each delta function is evaluated as
\begin{align}
\delta(\omega + |\bm{p} + \bm{q}| - |\bm{p}|)
 &= \frac{p-\omega}{pq}\delta\left( x - \frac{\omega^2 - 2p\omega - q^2}{2pq} \right)\theta(p-\omega), \\
\delta(\omega - |\bm{p} + \bm{q}| + |\bm{p}|)
 &= \frac{p+\omega}{pq}\delta\left( x - \frac{\omega^2 + 2p\omega - q^2}{2pq} \right)\theta(p+\omega), \\
\delta(\omega - |\bm{p} + \bm{q}| - |\bm{p}|)
 &= \frac{-p+\omega}{pq}\delta\left( x - \frac{\omega^2 - 2p\omega - q^2}{2pq} \right)\theta(-p+\omega), \\
\delta(\omega + |\bm{p} + \bm{q}| + |\bm{p}|)
 &= \frac{-p-\omega}{pq}\delta\left( x - \frac{\omega^2 + 2p\omega - q^2}{2pq} \right)\theta(-p-\omega),
\end{align}
where $x=\cos\theta$.
Considering the argument of the delta function, we find
\begin{align}
{\rm Im}\Pi_{\rm ps}^{0}(\bm{q}, \omega+i\eta)
 &= N_fN_c\frac{\omega^2 - q^2}{8\pi q} \int_0^\infty dp \bigg\{ \left[f(p + \mu) - f(p + \mu - \omega)\right] \theta(p-\omega)\theta\left(-1 < \frac{\omega^2 - 2p\omega - q^2}{2pq} < 1\right) \notag \\
 &~~~~~~~~~~~~~~~~~~ + \left[f(p - \mu + \omega) - f(p - \mu)\right] \theta(p+\omega) \theta\left(-1 < \frac{\omega^2 + 2p\omega - q^2}{2pq} < 1 \right) \notag \\
 &~~~~~~~~~~~~~~~~~~ - \left[- f(p + \mu - \omega) + f(p + \mu)\right] \theta(-p+\omega) \theta\left(-1< \frac{\omega^2 - 2p\omega - q^2}{2pq} < 1 \right) \notag \\
 &~~~~~~~~~~~~~~~~~~ - \left[f(p - \mu + \omega) - f(p - \mu)\right] \theta(-p-\omega) \theta\left(-1 < \frac{\omega^2 + 2p\omega - q^2}{2pq} < 1 \right) \bigg\} .
\end{align}
Therefore,
\begin{align}
{\rm Im}\Pi_{\rm ps}^{0}(\bm{q}, \omega+i\eta)
 &= N_fN_cT\frac{\omega^2 - q^2}{8\pi q} \Big\{ \ln\left[ 1 + e^{-\beta(q/2 + \mu + \omega/2)} \right] - \ln\left[ 1 + e^{-\beta(q/2 + \mu - \omega/2)} \right] \notag \\
 &~~~~~~~~~~~ + \ln\left[ 1 + e^{-\beta(q/2 - \mu + \omega/2)} \right] - \ln\left[ 1 + e^{-\beta(q/2 - \mu - \omega/2)} \right] + \beta\omega\theta(|\omega|-q) \Big\}.
\end{align}

\section{Integrals $I_n$}
\label{sec:c}
We evaluate the integrals $I_n$ by using the proper time formalism,
\begin{align}
I_n(r) = (-1)^{n-1} T\sum_m \int \frac{d^3\bm{q}}{(2\pi)^3}
 \int_0^\infty ds\,s^{n-1}
 e^{-s\left[r + \gamma \left( |\bm{q}|^2-q_c^2 \right)^2 + \alpha|\omega_m|\right]}.
\end{align}
The Matsubara frequency can be summed up as follows:
\begin{align}
I_n(r) = (-1)^{n-1} T \int \frac{d^3\bm{q}}{(2\pi)^3}
 \int_0^\infty ds\,s^{n-1}
 e^{-s\left[r + \gamma \left( |\bm{q}|^2-q_c^2 \right)^2 \right]}\coth(\alpha \pi Ts).
\end{align}
If $r$ is sufficiently small,
 the main region contributed to the integral is $|\bm{q}|\sim q_c$.
Therefore, the integral can be approximated as
\begin{align}
I_n(r) &\simeq \frac{(-1)^{n-1}T}{4\pi^2} \int_{-\infty}^\infty q^2dq \int_0^\infty ds\,s^{n-1}\left\{ e^{-s\left[r + 4\gamma q_c^2\left( q-q_c \right)^2 \right]} + e^{-s\left[r + 4\gamma q_c^2\left( q+q_c \right)^2 \right]} \right\}\coth(\alpha \pi Ts) \notag \\
 &= \frac{(-1)^{n-1}T}{2\pi^2} \int_0^\infty ds \left( \frac{1}{2} \sqrt{\frac{\pi}{(4\gamma q_c^2 s)^3}} + \sqrt{\frac{\pi q_c^2}{4\gamma s}}\right) s^{n-1}e^{-sr} \coth(a_1 \pi T s).
\end{align}
When $n=1,2$,
 the proper time regularization should be introduced
 because the integrals have an ultraviolet divergence.
However, the reading contribution is not affected
 by the cutoff at $r\sim0$, except for $n=1$ at $T=0$.

\section{Correlation function in the DCDW phase}
\label{sec:d}
In the case of $T=0$, the correlation function takes the form
\beqa
\langle \left[\beta_i(z) - \beta_i(0)\right]^2 \rangle_{T=0} &=& c_i\int \frac{d^4k}{(2\pi)^4}\frac{1-e^{ik_zz}}{k_0^2+A_ik_z^2+B_ik_\perp^4} \nonumber \\
&=& \frac{c_i}{16\pi^3\sqrt{B_i}} \int^\infty_{-\infty} dk_0 \int^\infty_{-\infty} dk_z \int^\infty_0 dx \frac{1-e^{ik_zz}}{k_0^2+A_ik_z^2+x^2},
\eeqa
where $x = \sqrt{B_i}k_\perp^2$.
Furthermore, putting $y=\sqrt{x^2+k_0^2}$, we can obtain
\beqa
\langle \left[\beta_i(z) - \beta_i(0)\right]^2 \rangle_{T=0} &=& \frac{c_i}{16\pi^2\sqrt{B_i}} \int^\infty_{-\infty} dk_z \int^\infty_0 ydy \frac{1-e^{ik_zz}}{y^2+A_ik_z^2} \nonumber \\
&=& \frac{c_i}{16\pi^2\sqrt{A_iB_i}} \int^\Lambda_0 dy \left( 1 - e^{-A^{-1/2}_iy|z|} \right) \nonumber \\
&\sim& \frac{c_i}{16\pi\sqrt{A_iB_i}}\Lambda~~~{\rm for~large}~|z|,
\eeqa
where the ultraviolet cutoff $\Lambda$ is inserted.
In the case of $T\neq0$, on the other hand,
 the leading contribution comes from the lowest Matsubara frequency,
\beqa
 \langle \left[\beta_i(z) - \beta_i(0)\right]^2 \rangle_{T\neq0} &\sim& c_iT\int \frac{d^3\bm{k}}{(2\pi)^3}\frac{1-e^{ik_zz}}{A_ik_z^2+B_ik_\perp^4} \nonumber \\
 &=& \frac{c_iT}{8\pi^2\sqrt{B_i}} \int^\infty_{-\infty} dk_z \int^\infty_0 dx \frac{1-e^{ik_zz}}{A_ik_z^2+x^2} \nonumber \\
 &=& \frac{c_iT}{8\pi\sqrt{A_iB_i}} \int^\Lambda_0 dx \frac{1-e^{-A_i^{-1/2}x|z|}}{x}.
\eeqa
Here we insert the convergence factor,
\beqa
 \langle \left[\beta_i(z) - \beta_i(0)\right]^2 \rangle_{T\neq0} &\sim& \lim_{\epsilon \to +0}\frac{c_iT}{8\pi\sqrt{A_iB_i}} \int^{A_i^{-1/2}\Lambda |z|}_0 dx \frac{1-e^{-x}}{x^{1-\epsilon}} \nonumber \\
 &=& \lim_{\epsilon \to +0}\frac{c_iT}{8\pi\sqrt{A_iB_i}} \left[ \frac{1}{\epsilon} \left(A_i^{-1/2}\Lambda |z| \right)^\epsilon - \Gamma(\epsilon) +\Gamma(\epsilon,A_i^{-1/2}\Lambda |z|)\right]\nonumber \\
&\sim& \frac{c_iT}{8\pi\sqrt{A_iB_i}} \ln \left( A_i^{-1/2}\Lambda |z| \right)~~~{\rm for~large}~|z|.
\eeqa


\begin{thebibliography}{99}

\bibitem{fuk}
K. Fukushima and T. Hatsuda, Rept. Prog. Phys. {\bf 74}, 014001 (2011);
K. Fukushima and C. Sasaki, Prog. Part. Nucl. Phys. {\bf 72}, 99 (2013).

\bibitem{chi}
D. V. Deryagin, D. Yu. Grigoriev, and V.A. Rubakov, Int. J. Mod. Phys. A {\bf 7}, 659 (1992);
E. Shuster and D. T. Son, Nucl. Phys. B {\bf 573}, 434 (2000);
B.-Y. Park, M. Rho, A. Wirzba, and I. Zahed, Phys. Rev. D {\bf 62}, 034015 (2000);
R. Rapp, E. Shuryak, and I. Zahed, Phys. Rev. D {\bf 63} 034008 (2001). 

\bibitem{dcdw}
T. Tatsumi and E. Nakano, hep-ph/0408294;
E. Nakano and T. Tatsumi, Phys. Rev. D {\bf 71}, 114006 (2005).

\bibitem{nic}
D. Nickel, Phys. Rev. Lett. {\bf 103}, 072301 (2009);
Phys. Rev. D {\bf 80}, 074025 (2009).

\bibitem{bub}
M. Buballa and S. Carignano, Prog. Part. Nucl. Phys. {\bf 81}, 39 (2015).

\bibitem{fro}
I. E. Frolov, V. Ch. Zhukovsky, and K. G. Klimenko, Phys. Rev. D {\bf 82}, 076002 (2010). 

\bibitem{tatb}
T. Tatsumi, K. Nishiyama, and S. Karasawa, Phys. Lett. B {\bf 743}, 66 (2015).

\bibitem{yos1}
R. Yoshiike, K. Nishiyama, and T. Tatsumi, Phys. Lett. B {\bf 751}, 123 (2015).

\bibitem{yos2}
R. Yoshiike and T. Tatsumi, Phys. Rev. D {\bf 92}, 116009 (2015).

\bibitem{lee}
T.-G. Lee, E. Nakano, Y. Tsue, T. Tatsumi, and B. Friman, Phys. Rev. D {\bf 92}, 034024 (2015).

\bibitem{hid}
Y. Hidaka, K. Kamikado, T. Kanazawa, and T. Noumi, Phys. Rev. D {\bf 92}, 034003 (2015).

\bibitem{lan}
L.D. Landau and E.M. Lifshitz, {\it Statistical Physics}, (Pergamon Press, Oxford, 1969).

\bibitem{pei}
R. E. Peierls, {\it Quantum Theory of Solids} (Oxford University Press, 1955). 

\bibitem{gen}
P. G. de Gennes and J. Prost, {\it The physics of liquid crystals} (Oxford University Press, 1974);
P. M. Chaikin and T. C. Lubensky, {\it Principles of condensed matter physics} (Cambridge Univ Press, 2000);
W. H. de Jeu, B. I. Ostrovskii, and A. N. Shalaginov, Rev. Mod. Phys. {\bf 75}, 181 (2003).

\bibitem{fflo}
H. Shimahara, J. Phys. Soc. Jpn. {\bf 67}, 1872 (1998);
H. Shimahara, Physica B: Condensed Matter {\bf 259}, 492 (1999).

\bibitem{fflo2}
L. Radzihovsky and A. Vishwanath, Phys. Rev. Lett. {\bf 103}, 010404 (2009);
L. Radzihovsky, Phys. Rev. A {\bf 84}, 023611 (2011);
 Physica C: Superconductivity {\bf 481}, 189 (2012).

\bibitem{gia}
T. Giamarchi and P. Le Doussal, Phys. Rev. Lett. {\bf 72}, 1530 (1994);
 Phys. Rev. B {\bf 52}, 1242 (1995);
T. Nattermann and S. Scheidl, Adv. Phys. {\bf 49}, 607 (2000).

\bibitem{bfg}
G. Baym, B. Friman, and G. Grinstein, Nucl. Phys. B {\bf 210}, 193 (1982).

\bibitem{bkt}
V. Berezinskii, Sov. Phys. JETP {\bf 32}, 493 (1971);
J. M. Kosterlitz and D. J. Thouless, J. Phys. C {\bf 6}, 1181 (1973).

\bibitem{had}
Z. Hadzibabic, P. Kr\"uger, M. Cheneau, B. Battelier, and J. Dalibard, Nature {\bf 441}, 1118 (2006).

\bibitem{mur}
P. A. Murthy, I. Boettcher, L. Bayha, M. Holzmann, D. Kedar, M. Neidig, M. G. Ries, A. N. Wenz, G. Zurn, and S. Jochim, Phys. Rev. Lett. {\bf 115}, 010401 (2015).

\bibitem{nit}
W. H. Nitsche, N. Y. Kim, G. Roumpos, C. Schneider, M. Kamp, S. H\"ofling, A. Forchel, and Y. Yamamoto, Phys. Rev. B {\bf 90}, 205430 (2014).

\bibitem{2d}
S. Carignano and M. Buballa, Phys. Rev. D {\bf 86}, 074018 (2012).

\bibitem{mig}
A. B. Migdal, Rev. Mod. Phys. {\bf 50} (1978) 107;
A .B. Migdal, E. E. Saperstein, M. A. Troitsky, and D. N. Voskresensky, Phys. Rep. {\bf 192}, 179 (1990).

\bibitem{super}
P. Fulde and R. A. Ferrel, Phys. Rev. {\bf 135}, A550 (1964);
A. I. Larkin and Y. N. Ovchinnikov, Sov. Phys. JETP {\bf 20}, 762 (1965);
R. Casalbuoni and G. Nardulli, Rev. Mod. Phys. {\bf 76}, 263 (2004);
Y. Liao, A. S. C. Rittner, T. Paprotta, W. Li, G. B. Partridge, R. G. Hulet, S. K. Baur, and E. J. Mueller, Nature {\bf 467}, 567 (2010).


\bibitem{bra}
S. A. Brazovskii, Sov. Phys. JETP {\bf 41}, 85 (1975).

\bibitem{dyu}
A. M. Dyugaev, Sov. Phys. JETP Lett. {\bf 22}, 83 (1975).

\bibitem{hoh}
P. C. Hohenberg and J. B. Swift, Phys. Rev. E {\bf 52}, 1828 (1995).

\bibitem{fre}
G. H. Fredkickson and K. Binder, J. Chem.Phys. {\bf 91}, 7265 (1989).

\bibitem{oha}
Y. Ohashi, J. Phys. Soc. Jpn. {\bf 71}, 2625 (2002).

\bibitem{nsr}
P. Nozi\`eres and S. Schmitt-Rink, J. Low Temp. Phys. {\bf 59}, 195 (1985).

\bibitem{gri}
Y. Ohashi and A. Griffin, Phys. Rev. Lett. {\bf 89}, 130402 (2002);
 Phys. Rev. A {\bf 67}, 063612 (2003).

\bibitem{lok}
V. M. Loktev, R. M. Quick, and S. G. Sharapov, Phys. Rep. {\bf 349}, 1 (2001).

\bibitem{kit}
M. Kitazawa, T. Koide, T. Kunihiro, and Y. Nemoto, Phys. Rev. D {\bf 65} (2002) 091504;
 {\bf 70}, 056003 (2004).
 
\bibitem{lin}
D. D. Ling, B. Friman, and G. Grinstein, Phys. Rev. B {\bf 24}, 2718 (1981).

\bibitem{lei}
L. Leibler, Macromolecules {\bf 13}, 1602 (1980).

\bibitem{bat}
F. S. Bates, J. H. Rosedale, and G. H. Fredickson, J. Chem. Phys. {\bf 92}, 6255 (1990).

\bibitem{kar}
S. Karasawa, T.-G. Lee, and T. Tatsumi, Prog. Theor. Exp. Phys. {\bf 2016}, 043D02 (2016).

\bibitem{hal}
B. I. Halperin, T. C. Lubensky, and S. Ma, Phys. Rev. Lett. {\bf 32}, 292 (1974).

\bibitem{bak}
P. Bak, S. Krinsky, and D. Mukamel, Phys. Rev. Lett. {\bf 36}, 52 (1976).

\bibitem{bin}
K. Binder, Rep. Prog. Phys. {\bf 50}, 783 (1997).

\bibitem{jan}
M. Janoschek, M. Garst, A. Bauer, P. Krautscheid, R. Georgii, P. B\"{o}ni, and C. Pfleiderer, Phys. Rev. B {\bf 87}, 134407 (2013).

\bibitem{tho}
D. J. Thouless, Ann. Phys. {\bf 10}, 553 (1960).

\bibitem{abr}
A. A. Abrikosov, L. P. Gorkov, and I. E. Dzyaloshinskii, {\it Methods of Quantum Field Theory in Statistical Physics} (Prentice-Hall, Inc., Englewood Cliffs, New Jersey, 1963).

\bibitem{fet}
A. L. Fetter and J. L. Walecka, {\it Quantum Theory of Many-Particle Systems} (McGraw-Hill, , New York, 1971).

\bibitem{kle}
S. P. Klevansky, Rev. Mod. Phys. {\bf 64}, 649 (1992).




\bibitem{cmw}
S. R. Coleman, Commun. Math. Phys. {\bf 31}, 259 (1973);
N. D. Mermin and H. Wagner, Phys. Rev. Lett. {\bf 17}, 1133(1966);
P. C. Hohenberg, Phys. Rev. {\bf 158}, 383 (1967).

\bibitem{wen}
X.-G. Wen, {\it Quantum Field Theory of Many-Body Systems} (Oxford University Press, 2004).

\bibitem{klei}
H. Kleinert, Phys. Lett. B {\bf 102}, 1 (1981).

\bibitem{rad}
L. Radzihovsky, Phys. Rev. {\bf A84}, 023611 (2011).



\bibitem{fuj}
H. Fujii, Phys. Rev. D {\bf 67}, 094018 (2003);
H. Fujii and M. Ohtani, Phys. Rev. D {\bf 70}, 014016 (2004).

\bibitem{sha}
R. Shankar, Rev. Mod. Phys. {\bf 66}, 129 (1994).

\bibitem{car}
S. Carignano, D. Nickel and M. Buballa, Phys. Rev. D {\bf 82}, 054009 (2010).

\bibitem{lee2}
T.-G. Lee, R. Yoshiike and T. Tatsumi, in preparation.

\end{thebibliography}
\end{document}